\documentclass{sig-alternate-2013}
\newfont{\mycrnotice}{ptmr8t at 7pt}
\newfont{\myconfname}{ptmri8t at 7pt}
\let\crnotice\mycrnotice%
\let\confname\myconfname%

\permission{Permission to make digital or hard copies of all or part of this work for personal or classroom use is granted without fee provided that copies are not made or distributed for profit or commercial advantage and that copies bear this notice and the full citation on the first page. Copyrights for components of this work owned by others than ACM must be honored. Abstracting with credit is permitted. To copy otherwise, or republish, to post on servers or to redistribute to lists, requires prior specific permission and/or a fee. Request permissions from Permissions@acm.org.}
\conferenceinfo{SIGIR'15,}{August 09 - 13, 2015, Santiago, Chile.}
\copyrightetc{\copyright~2015 ACM. ISBN \the\acmcopyr}
\crdata{978-1-4503-3621-5/15/08\ ...\$15.00.\\
DOI: http://dx.doi.org/10.1145/2766462.2767717}

\usepackage{multirow}
\usepackage{color}
\usepackage{algorithm}
\usepackage{algorithmicx}
\usepackage{algpseudocode}
\usepackage{colortbl}
\usepackage{pgf}
\usepackage{tikz}
\usepackage{pgfplots}
\usepackage{todonotes}
\usepackage{footnote}

\begin{document}
%

\title{Non-Compositional Term Dependence \\for Information Retrieval}
%
%
%
%
%

\numberofauthors{2} 
%
\author{
%
%
\alignauthor
Christina Lioma\\
       \affaddr{Department of Computer Science}\\
       \affaddr{University of Copenhagen, Denmark}\\
       \email{c.lioma@di.ku.dk}
\alignauthor
Jakob Grue Simonsen\\
       \affaddr{Department of Computer Science}\\
       \affaddr{University of Copenhagen, Denmark}\\
       \email{simonsen@di.ku.dk}
\and
\alignauthor
Birger Larsen\\
       \affaddr{Department of Communication}\\
       \affaddr{Aalborg University Copenhagen, Denmark}\\
       \email{birger@hum.aau.dk}
\alignauthor
Niels Dalum Hansen\\
       \affaddr{Department of Computer Science}\\
       \affaddr{University of Copenhagen, Denmark}\\
       \email{nhansen@di.ku.dk}
}

\date{30 July 1999}

\maketitle
\begin{abstract}
Modelling term dependence in IR aims to identify co-occur-\\ring terms that are too heavily dependent on each other to be treated as a bag of words, and to adapt the indexing and ranking accordingly. Dependent terms are predominantly identified using lexical frequency statistics, assuming that (a) if terms co-occur often enough in some corpus, they are semantically dependent; 
(b) the more often they co-occur, the more semantically dependent they are. This assumption is not always correct: the frequency of co-occurring terms can be separate from the strength of their semantic dependence. E.g. \texttt{red tape} might be overall less frequent than \texttt{tape measure} in some corpus, but this does not mean that \texttt{red+tape} are less dependent than \texttt{tape+measure}. This is especially the case for \textit{non-composi\-tional phrases}, i.e. phrases whose meaning cannot be composed from the individual meanings of their terms (such as the phrase \texttt{red tape} meaning bureaucracy). 

Motivated by this lack of distinction between the frequency and strength of term dependence in IR, we present a principled approach for handling term dependence in queries, using both lexical frequency and semantic evidence. We focus on non-compositional phrases, extending a recent unsupervised model for their detection \cite{KielaC13} to IR. Our approach, integrated into ranking using Markov Random Fields \cite{Metzler2005}, yields effectiveness gains over competitive TREC baselines, showing that there is still room for improvement in the very well-studied area of term dependence in IR.
\end{abstract}



\section{Introduction}
\label{s:intro}
Frege's \textit{principle of compositionality} posits that the meaning of an expression is a function of the meanings of its constituent expressions and the ways they combine \cite{frege}. Applied to linguistics by Montague, this principle implies that the meaning of some text is not just the collective meaning of its words, but also a function of how these words are arranged. Whereas this holds most of the times, occasionally language is \textit{non-compositional}, i.e. the meaning and arrangement of words alone is not enough to convey the overall semantics. E.g. the phrase \texttt{red tape} (meaning bureaucracy) is not a \texttt{tape} of type \texttt{red}. This linguistic phenomenon is known as \textit{non-compositionality}. 

The challenges posed by non-compositionality have spur-\\red Natural Language Processing (NLP) research in automatic non-composi\-tionality detection, e.g. in nouns \cite{Baldwin:2003,Schulte:2013}, verb-noun \cite{Katz:automatic} and verb-particle \cite{McCarthy:2003} combinations, using techniques such as latent semantic analysis \cite{Katz:automatic}, compositional translations to multiple languages \cite{Salehi:2013}, sense induction \cite{Korkontzelos:2009:DCM:1667583.1667605} and word space models \cite{krvcmavr-jevzek-pecina:2013:CVSC,McCarthy:2011}. 
An active line of research focuses on distributional and vector-based models of word and phrase meaning leading to vector-space models for compositionality \cite{LNC3:LNC3362,COGS:COGS1106,DBLP:conf/ijcnlp/ReddyKMM11}. These advances have not penetrated IR research notably (with the exception of \cite{mich11}, discussed in Section \ref{s:relw}), despite long and persistent IR interest in term dependence. A resulting risk is that the strength of term dependence may be consistently miscalculated in IR. We explain this next.

In IR, dependent terms are predominantly identified using lexical frequency statistics: if terms co-occur often enough in some typically large dataset, they are  assumed to be dependent, and the strength of their dependence is typically assumed proportional to their frequency of co-occurrence, e.g. see \cite{Fagan1989,MishneR2005}. Simply stated, the more frequently two terms co-occur, the more dependent we assume they are. This assumption is not always correct. In linguistics, the frequency of term co-occurrence can be somewhat separate from the strength of semantic dependence. Even though the former can be indicative to some extent of the latter, their relation is not symmetric. 
E.g. \texttt{red tape} might be overall less frequent than \texttt{tape measure} in some corpus, but this does not mean that \texttt{red+tape} are semantically less dependent than \texttt{tape+measure}; quite  the contrary. Non-compositionality lies at the heart of this because non-compositional terms are maximally 
 dependent, \emph{regardless} of their frequency of co-occurrence. So, whereas the strength of term dependence within compositional phrases, e.g., \texttt{tape measure, white horse}, can be reasonably approximated by their frequency of co-occurrence in a corpus, this is \textit{not} true for non-compositional phrases, like \texttt{red tape, dark horse}.

Motivated by this lack of distinction in IR between frequency of term co-occurrence and strength of term dependence, we present a principled approach for treating term dependence in queries. This approach extends a recent unsupervised model for detecting non-compositional phrases using lexical frequency and semantic evidence \cite{KielaC13}. 
The main idea consists of (a) substituting a term in a phrase by a synonym (e.g. \texttt{red tape} would become \texttt{scarlet tape}) and (b) measuring the semantic divergence of the replacement phrase from the original phrase. If their meanings diverge, the original phrase is more likely to be non-compositional. If however their meanings do not diverge much (e.g. \texttt{tax office} would become \texttt{tax bureau}), then the original phrase is less likely to be non-compositional. We extend the vector space model proposed for measuring this divergence in \cite{KielaC13} with a probabilistic model that measures the Kullback-Leibler divergence between the language models of the original and replacement phrase (Sections \ref{s:model}-\ref{s:ground}). 
We apply both approaches to detect strongly dependent query terms, which we then treat in a non-bag of words fashion during ranking (Section \ref{s:Evaluation}). Experiments with 350 TREC queries show that our approaches consistently outperform competitive baselines, and are particularly effective for 2-, 3-, and 4-term queries in the web search task.

\section{Related Work}
\label{s:relw}
Broadly speaking, efforts to model term dependence, also known as \textit{term co-occurrence, adjacency} and \textit{lexical affinities}\footnote{\textit{Dependence, co-occurrence, adjacency} and \textit{lexical affinities} are not synonyms \cite{Hey2002}, but in IR they are used interchangeably.} in IR, typically model phrases found in queries and/or documents, motivated by the intuition to consider as \emph{more} relevant those documents in which terms appear in the same order and patterns as they appear in the query, and as \emph{less} relevant those documents in which terms are separated \cite{SmithD1985}. These efforts were initiated mainly in the 1980s, and intensified in the 1990s, reporting retrieval benefits. Later, efforts decreased: baseline performance improved, and the cost associated with linguistic processing was not worth the small benefits over the already improved baselines  \cite{jones2005}. 

Generally, term dependence is detected using either statistical or linguistic information. Research began with the early work on statistical term associations \cite{Doyle1962, Giuliano1963, Lesk1969, Stiles1961} and syntax-based approaches \cite{Baxendale1958, Earl1972, Salton1966}, continuing with work on probabilistic term dependence models \cite{HarperK1978, Salton1982, Turtle1991, K1977, Yu1983}, syntactic methods \cite{DillonG1983, Metzler1984, SmeatonK1988, Smeaton1986} and statistical approaches \cite{Fagan1989, Lewis1992, LewisC1990}. From the mid-1990s onwards, research focused on hybrid methods combining syntactic and statistical approaches of phrase processing \cite{EvansZ1996}, phrase-based enhancement of the indexed term representations \cite{Zhai1997}, and phrase-based term weighting \cite{NaritaO2000,Pederson1997,Str1997}. More recent research has focused on statistical methods, primarily using language modelling \cite{Metzler2005, MishneR2005, NallapatiA2002, SongC1999, SrikanthS03} but not exclusively \cite{Losee1994, PlachourasO2007}, while attention has also been given to term dependence and efficient large-scale indexing \cite{Fujita2001, Lin2001}. The Markov Random Field (MRF) model of term dependence \cite{Metzler2005} reported significant improvements in retrieval effectiveness. 

Several more recent studies address term dependence, for instance using heuristics \cite{TaoZ07}, formalising the term position in the document \cite{LvZ09}, or extending the MRF model to concepts \cite{BenderskyMC10}, all reporting positive findings. This continued interest in term dependence may indicate that it is still an open problem. However, to our knowledge, none of these approaches addresses non-compositionality, except Michelbacher et al. \cite{mich11}, who focus primarily on the automatic detection of the head modifier inside non-compositional phrases and use IR as a task illustrating that the information they detect can be useful. They experiment with a small non-TREC dataset and report statistically significant gains in retrieval precision. 

\section{Non-Compositionality \\Detection (NCD)}
\label{s:model}

Non-Compositionality Detection (NCD) aims to identify the presence and strength of non-compositional phrases in language. This is typically realised as a measurement, i.e. through some function that outputs a compositionality score for a phrase. Given a scale of such scores, the minimum and maximum reflect the total absence of compositionality (non-compositionality), e.g. \texttt{red tape}, and complete compositionality, e.g. \texttt{tax office}, respectively. 
Sliding along such a scale corresponds to moving across phrases of various levels of compositionality, practically facilitating the comparison of phrases on the grounds of their term dependence. 
We reason that such a comparison may be useful to IR, where systems need to process differently queries at different positions of this scale, i.e. keyword-based (= compositional) queries such as \texttt{London transportation}, and queries containing heavily dependent terms (=non-compositional) such as \texttt{red tape AL register car}. 


This section presents how we use non-compositionality to model term dependence in IR. Among the various NCD approaches outlined in Section \ref{s:intro}, we use the recent approach of \cite{KielaC13} because it is unsupervised, resource-efficient, and performs competitively on benchmark tests. 
We extend this NCD approach, which uses vector spaces, by adding a second estimation of non-compositionality, this time probabilistic. In addition, we formally express the methodological description of \cite{KielaC13}  as a model of query perturbation, and we model non-compositional term dependence specifically for IR queries, not general phrases like \cite{KielaC13}, with considerations to data constraints in an IR context.  

\subsection{Non-compositionality in queries}
\label{ss:noncomp}

Given a query, we aim to detect the presence and strength of non-compositionality in it. 
Kiela and Clark \cite{KielaC13} posit that non-compositional phrases can be identified by substituting each of the original words in them, one at a time, by some other relevant/synonymous term, and comparing the meaning of each phrase resulting from the substitution to the meaning of the original phrase. The more they diverge, the less compositional the original phrase is. E.g. replacing \texttt{car} by \texttt{vehicle} in \texttt{import car} gives \texttt{import vehicle}, which is semantically similar to the original, but replacing \texttt{red} by \texttt{scarlet} in \texttt{red tape} gives \texttt{scarlet tape}, which is semantically different from the original. Hence, \texttt{red tape} is less compositional than \texttt{import car}. The core idea is that such substitutions are likely to have a low impact on the semantics of compositional phrases, but a high impact on the semantics of non-compositional phrases. The resulting \emph{semantic divergence} is then approximately inversely proportional to compositionality. 

Conceptually, we see this approach as applying perturbations over some signal in order to study the resulting effects upon the signal. In our case, the signal is the query and the perturbation is the replacement of a query term by another term. We express this perturbation as follows: 
Let $S_q(I;T)$ be the semantic space $S$ of query $q$ containing the ordered set of terms $T$, where $I$ is the information conveyed by $T$. 
Let $\bar{T}$ denote the ordered set of query terms where one of them has been replaced by another term (e.g., a synonym). I.e., 
$\bar{T}$ is the perturbation of $T$. 
Then, the non-compositionality $N_q$, and the compositionality $C_q$ of query $q$, can be expressed as a function of the divergence $\psi$ of the resulting semantic spaces $\psi(S_q(I;T), S_q(I;\bar{T}))$, for all $m = \vert T \vert$ divergences resulting from all substitutions:
\begin{align}
\label{eq:delta}
N_q &= f\{\psi(S_q(I;T),S_q(I;\bar{T}) : \bar{T} \in \{T_1,\ldots,T_m\})\} \\
 C_q &= g(N_q)
\end{align}
\noindent where $f$ is typically some summation or averaging function over the set of divergences, and $g$ is some decreasing function, e.g. $g(x) \mapsto 1/x$ . Thus, non-compositionality increases with semantic divergence,  but compositionality \emph{de}creases with semantic divergence. 

Unless constrained, such perturbations risk drifting semantically further away than intended, e.g., if two out of all three terms in a query are replaced simultaneously. Kiela \& Clarke address this using two constraints, which we also adopt: (a) only one  term is replaced at a time, and (b) a term is replaced by its synonym or a closely related term such as a hyper- or hyponym\footnote{We henceforth refer to all forms of closely related terms as synonyms.}. 
We identify a further risk of degrading performance by considering too many synonyms:
The set of perturbations for query $q$ consisting of terms $t_1 \cdots t_m$, where $s_j$  is a synonym of $t_j$, is:
$$
\{p_1,\ldots,p_m\} = \left\{ 
\begin{array}{c}
s_1 t_2 \cdots t_m\\
t_1 s_2 \cdots t_m \\
\vdots\\
t_1 \cdots t_{j-1} s_j t_{j+1} \cdots t_m \\
\vdots \\
t_1 \cdots t_{m-1} s_m
\end{array}
\right\}
$$
where $p_m$ denotes the $m^{th}$ perturbation. 
As a term may have more than one synonym, of various grades of synonymity, the set of perturbations can grow to include all synonyms of the query terms. A selection process must control the perturbations so that: (a) ``best possible" synonyms (as opposed to near-synonyms) are used, and (b) the number of perturbations is minimised, i.e. we perturb the queries no more than necessary for computing their compositionality. We thus use one perturbation per query term and experimentally show (in Section \ref{s:Evaluation}) that this suffices for IR. Other tasks, including NCD \textit{per se}, may require more perturbations per term.

\subsection{Semantic divergence}
\label{ss:diverg}
Computing the divergence in Equation \ref{eq:delta} requires that both the query and perturbations  be represented in some semantic space that is tractable and amenable to measurement. Kiela \& Clarke propose vector spaces (Section \ref{sss:vs}). We propose probability spaces as a complementary representation (Section \ref{sss:prob}). We present and experiment with both.

\subsubsection{Vector Space}
\label{sss:vs}
We re-express the vector space representation of Kiela \& Clarke for queries and their perturbations as follows. 
Let $\vec{v}(q)$ and $\vec{v}(p_j)$ be the vector of query $q$ and its perturbation $p_j$ respectively. The semantic divergence $\psi$ between the query and its perturbation can be modelled as the distance $d$ between their vectors ($\psi \approx d$), where $d$ is some appropriate distance function. This $d$ can be chosen as any vector distance function, e.g. Euclidean, Chebychev, or the better-known Cosine we use here. Then, assuming a summation function $f$ in Equation \ref{eq:delta}, the non-compositionality of a query containing terms $t_1 \cdots t_m$ of $k$ synonyms is:
\begin{eqnarray}
\frac{1}{mk} \sum_{j=1}^m \sum_{i=1}^k d( \vec{v}(q),\vec{v}(p_{ij}) )
\end{eqnarray}
where $p_{ij}$ is the perturbation $t_1 \cdots t_{j-1} s_{ij} t_{j+1} \cdots t_m$ and $s_{ij}$ is the $i^{th}$ synonym of term $t_j$. Using one synonym per term only (as we do) reduces this to:
\begin{eqnarray}
\label{eq:dist}
\frac{1}{m} \sum_{j=1}^m d({\vec{v}(q)},{\vec{v}(p_j)})
\end{eqnarray}


%
%

The main idea is to represent a query and its perturbation as vectors, so that we can interpret their distance as semantic divergence. Practically this means mapping $\psi$ from Equation\ (\ref{eq:delta}) to $d$ above.
Dating back to Salton, the IR and NLP literature abounds with variations of how the above vector representation can be implemented and interpreted, any of which can be used here. We describe how we build the vectors and how we compare their distance in Section \ref{s:ground}.

\subsubsection{Kullback-Leibler Divergence}
\label{sss:prob}
We now present our probabilistic representation of queries and their perturbations. The high-level difference from the previous representation is that instead of representing a query as a vector of terms, we represent it as a distribution of events, where the events correspond to terms. Such representations are called probabilistic because they allow computing the probability of an event occurring, i.e. the probability of a term occurring in the query. When these probabilities are interpreted in a frequentist way, they are approximated by relative frequencies (i.e. normalised word counts). In text processing, this is known as language modelling.

We reason that, if queries and their perturbations are represented as event distributions, then their divergence can be computed using standard methods, one of the better known being their Kullback-Leibler divergence (KLD). 
Even though KLD is not a distance metric (it is not symmetric), it is widely used in IR to approximate the semantic distance between texts, where higher KLD values indicate more divergence.
We apply this to compute the semantic divergence $\psi$ in Equation \ref{eq:delta}, by building a language model for the query and each perturbation. Then, their KLD should be proportional to the semantic divergence $\psi$ in Equation\ (\ref{eq:delta}), i.e. ($\psi \approx KLD$). Let $LM_q$ and $LM_p$ denote the language models of query $q$ and perturbation $p$ respectively. Their KLD is:

\begin{equation}
\label{eq:kld}
KLD(LM_q || LM_p) = LM_q \log{\frac{LM_q}{LM_p}}
\end{equation}

We next describe how we build $LM_q,LM_p$ and how we operationalise Equation\ (\ref{eq:kld}).

\section{Model induction}
\label{s:ground}
Both vector and probability space representations presented above approximate how different a perturbation is from the original query, albeit in different ways. This section describes their exact mechanics.

We start by describing what the above vectors and language models actually consist of. As the approach is the same for both queries and their perturbations, we henceforth refer to their union as $Q$. For each term $t \in Q$, we build a \textit{context window} as follows: we extract bags of terms occurring within a window of maximum $n$ terms away from $t$ in some large document corpus, so that the window consists of $2n+1$ terms. E.g., if $n$=5, then we consider 11 terms in total: 5 (immediately preceeding $t$) + $t$ + 5 (immediately succeeding $t$). The underlying assumption is that all the terms in a document have some relationship to all other terms in the document, modulo window size, outside of which the relationship is not taken into consideration. In statistical NLP this is a standard way of inducing word semantics from ``the company they keep", a.k.a. distributional semantics \cite{firth}. 
These context windows provide the ingredients of the vector and probabilistic representation of our queries and perturbations, explained next. 

\subsection{Vector representation}
\label{ss:vector}

After all context windows of a term $t \in Q$ are extracted, we compute a term weight vector $w_{t}$ for $t$ with the aim of capturing the salience of term $t$. Kiela \& Clarke show that such weights can function in a discriminative way for the task of NCD. For each query, we generate a term weight vector by combining the term weight vectors of the terms in the query. Next we explain how we compute the weights of the individual query terms and the weight of the whole query or perturbation.

\subsubsection{Individual Term Weights}
Kiela \& Clarke experiment with these five well-known weighting schemes, adapted to the context window scenario, (even though they only report results from LTU), which we also use:



\begin{center}
\bf ATC \cite{Reed2006}:
\end{center}
\begin{equation}
w_{it} = \frac{\left(0.5 + 0.5 \times \frac{f_{it}}{max_f}) \log (\frac{N}{n(t)}\right)}{ \sqrt{\sum^N_{i=1} \left( \left(0.5 + 0.5 \times \frac{f_{it}}{max_f}\right) \log \left(\frac{N}{n(t)}\right) \right)^2 }}
\end{equation}

\begin{center}
\bf LTU \cite{Singhal97}:
\end{center}
\begin{equation}
w_{it} = \frac{  (\log(f_{it}) + 1.0) \log \frac{N}{n(t)}   } {   0.8 + 0.2 \frac{M_i} { av.M}   }  
\end{equation}

\begin{center}
\bf Mutual Information (MI) \cite{PantelL02}:
\end{center}
\begin{equation}
w_{it} = \log \frac{\frac{f_{jt}}{N}}{\frac{\sum^N_{j=1}f_{jt}}{N} \times \frac{\sum^{M_i}_{k=1}f_{ik}}{N}} 
\end{equation}

\begin{center}
\bf Okapi \cite{JinFH01}:
\end{center}
\begin{equation}
w_{it} = \left(\frac{f_{it}}{0.5 + 1.5 \times \frac{ M_i } { av.M } + f_{it} }\right) \times \log \left( \frac{N - n(t) + 0.5}{f_{it} + 0.5}\right)
\end{equation}

\begin{center}
\bf  TFxIDF \cite{ksj72}:
\end{center}
\begin{equation}
w_{it} = \log(f_{it}) \times  \log\left(\frac{N}{n(t)}\right) 
\end{equation}

\noindent where $w_{it}$ is the weight of term $t$ in context window $i$; 
$f_{it}$ is the frequency of $t$ in context window $i$;
$N$ is the total number of context windows;
$n(t)$ is the number of context windows containing $t$;
$M_i$ is the number of terms in context window $i$;
$av.M$ is the  average number of terms in all context windows; 
and $max_f$ is the maximum frequency of any term in any context window.

 To construct a vector $\vec{v}(t)$ for each $t \in Q$, we extract the context windows for $t$, which we denote $cw_t$. For each term, $t'$, represented by an entry in $\vec{v}(t)$, the corresponding weight is computed as the average of  $w_{it'}$ for $i \in cw_t$.

\subsubsection{Query/Perturbation Weights}
Having built such a vector for each $t \in Q$, the vector of the entire query or perturbation can be constructed in several ways, for instance as the element-wise sum of the vectors of its terms, or as their dilation, or as their pointwise multiplication. 
We choose the latter because it has been shown more effective for semantic vector representations in NLP \cite{KielaC13,COGS:COGS1106}. 
The final query vector $\vec{q}$ for query $q$ consisting of terms $t_1 \cdots t_m$ is:
\begin{equation}
\vec{v}(q) = \vec{v}(t_1) \odot \cdots \odot \vec{v}(t_m)
\end{equation}
where $\odot$ is the binary operator on equal-length vectors of real numbers defined by $(x_1,\ldots,x_n) \odot (y_1,\ldots,y_n) = (x_1 \times y_1, \ldots, x_n \times y_n)$. The perturbation vectors are built identically to this. 
Note
that as $\odot$ is associative and commutative, the $j^{th}$ component of $\vec{v}(q)$ is simply the product of all the $j^{th}$ components of
the vectors $\vec{v}(t_1), \ldots, \vec{v}(t_m)$.

As Kiela \& Clarke point out, using pointwise multiplication has a somewhat `reverse' effect on the semantic distance: overlapping components (i.e. terms appearing in common contexts) are stressed; since their vectors have little overlap outside the non-compositional meaning, their perturbations also have little overlap, resulting in a smaller change in distance when perturbed.  Another effect of pointwise multiplication is that the frequency of terms occurring in the context windows of a query term will be strengthened: if a term
$t$ has a high weight in both $\vec{v}(t)$ and $\vec{v}(t')$, it will have a high weight in $\vec{v}(t) \odot \vec{v}(t')$; however, low weight in either one of $\vec{v}(t)$
or $\vec{v}(t')$ will correspond to low weight in $\vec{v}(t) \odot \vec{v}(t')$. This means that the vectors of the terms of non-compositional queries, which will in general occur
in very different contexts, will have entries with fairly low absolute values. 
In contrast, for compositional queries, substituting a term by its synonym may yield constructions that can be expected to occur in a number of contexts
wildly different from the original, hence will have markedly different contextual statistics and thus greater distance $d$.


\subsection{Language modelling representation}
\label{ss:lm}
The alternative representation we propose for queries and perturbations is to use the set of all context windows of the terms in a query or perturbation to build a respective language model $LM_q, LM_p$ (introduced in Equation \ref{eq:kld}). 
There exist various ways of building language models from term counts, involving some sort of smoothing of the counts; we use two among the best known, \textit{Laplace} and \textit{Simple Good-Turing}.

\textit{Laplace} (or add-one) estimates the probability of a term $t$ in the language model of query $q$, $P_{LP}(q,t)$, as: 
\begin{equation}
\label{eq:lp}
P_{LP}(q,t)=\frac{c_{q,t} + 1}{C_{q}+V}
\end{equation}

\noindent where 
$c_{q,t}$ is the count of $t$ in $q$, $C_{q}$ is the count of all terms in the context windows of $q$, and $V$ is the number of terms in the language model of $q$. We compute it identically for perturbations (replacing $_q$ by $_p$ above). 

For sparse data over large vocabularies, Laplace tends to make a very big change to the counts and resulting probabilities because it moves too much probability mass to all unseen events (zero counts). We could move a bit less mass by adding a fractional count rather than 1 (e.g. add ``$\delta$-smoothing'' \cite{Jeffreys48}), but that would require choosing $\delta$ dynamically, risking inappropriate discounting for many counts, and producing overall counts with poor variances \cite{JurafskyM09}. For these reasons, we also apply \textit{Simple Good-Turing} \cite{gale1995good} smoothing, which uses (i) the counts of \textit{hapax legomena} (events occurring once) to estimate the counts of unseen events, and (ii) \textit{double counts}, i.e. the frequency of a frequency. 
Simple Good-Turing estimates the probability of a term $t$ with frequency $r$ in the language model of query $q$, $P_{GT}(q,t)$, as:

\begin{equation}
P_{GT}(q,t) = 
\frac{(r + 1) \cdot S(ff_{r+1})}{C_q \cdot S(ff_r)}  \;\;\mbox{for} \;\; r>0
\end{equation}
\noindent where
$ff$ is a vector with frequencies for term frequencies, $C_q$ is as defined as in Equation \ref{eq:lp}, and $S$ is a function fitted through the observed values of $ff$ to get the \textit{expected} count of these values (see \cite{gale1995good} for more). For zero count values the probability is calculated as follows:

\begin{equation}
\label{eq:gt}
P_{GT}(q,t) = 
\frac{ff_1}{C_q} \;\;\mbox{for} \;\; r=0
\end{equation}
\noindent where $ff_1$ is the frequency of frequency of \textit{hapax legomena}. We normalise the resulting language model to sum to 1. Simple Good-Turing is known to perform well, especially for large numbers of observations drawn from large vocabularies.

The above two smoothing methods produce a language model for \textit{each term} per query or perturbation. To produce one language model for the whole query or perturbation, we sort the language models of their terms and combine them in four different ways: (1) summing their values in quantiles 2 \& 3\footnote{We use quantiles 2 \& 3 to avoid outliers.}; (2) averaging their values in quantiles 2 \& 3; (3) multiplying their values; (4) using the median of their values. 
Overall, the above 2 smoothing methods $\times$ 4 combinations produce 8 language modelling variations of NCD.

\section{Discussion of our NCD approach}
Both representations (vector and probability space) of the NCD approach we present are parameterised over the notion of semantic divergence, which we operationalise with different weightings, each corresponding to some variation of computing this divergence. Our use of semantic divergence, measured typically as a real number in the model, corresponds to the observation that compositionality is not dichotomous: phrases in general are not only compositional or
non-compo\-sitional; rather, a fine-grained range of compositionality exists, a fact corroborated by human raters asked to score degrees of compositionality \cite{bannard2003statistical,McCarthy:2003}. Suitable divergence functions that could mimic the scores of human raters may exist, but we have not attempted to do so.

We have also not attempted to estimate the semantic `accuracy' of the phrases resulting from each perturbation, i.e. the extent to which they are non-sensical, even though Kiela \& Clarke state that this is possible with their approach \cite{KielaC13}. We estimate solely the divergence between the query and a perturbation, and not how much sense the perturbed phrase makes, for two reasons: (a) we reason that the semantic divergence should in principle suffice for indicating compositionality as we intend to use it in IR; (b) to our knowledge, no scalable automatic approach can adequately approximate such a semantic assessment for query logs. 

Another point of departure from Kiela \& Clarke is our treatment of query terms as a list, i.e. a set endowed with a strict order.
In principle, all computations presented, both by Kiela \& Clarke and by us, can be used with ordinary (i.e., unordered) sets of terms too, as has also been done with term dependence models in IR \cite{Metzler2005}. We use strictly ordered sets because non-compositionality is never manifested in language in any other way, for instance by mixing the order of non-compositional terms, or by interrupting them by another term. E.g., \texttt{red tape} can function non-compositonally (and mean bureaucracy) only when the terms \texttt{red} and \texttt{tape} appear adjacent and in that specific order. Ergo, no variation of \texttt{red .+ tape} or \texttt{tape .+ red} (in RegEx notation) can have the non-compositional meaning of bureaucracy. 


Finally, perturbations are common in science, and the practice of perturbing queries has even been used in IR before, albeit for different reasons. For instance Vinay et al.\ \cite{VinayCMW06} employ different query (and document) perturbations for query performance prediction: by altering the query term weights, they observe the documents retrieved, and study the relationship between the amount, or \textit{sensitivity of perturbation} and the quality of the ranking. Our approach, apart from having a different overall scope, namely  term dependence as opposed to query performance prediction, also differs from \cite{VinayCMW06} in that it applies a linguistically informed selection process for each perturbation: we replace query terms by their synonyms, not by varying their respective term weights within some range.

\section{Evaluation}
\label{s:Evaluation}

\subsection{Using NCD for selective term dependence}
This section presents experiments aiming to quantify the effectiveness of processing query term dependence, not as a bag of words, but as a `set phrase' of strict ordered adjacency, i.e. matching documents that contain an identical (ordered \& uninterrupted) sequence of terms. The main idea is to use NCD to select which among a batch or stream of queries contain dependent terms, and process only those queries as a `set phrase'; the rest of the queries can be processed as a bag of words.
For this initial study, we focus on the non-compositionality of the \textit{whole query}, not of phrases \textit{within} queries. 

We use the non-compositionality score of each query (computed with any of the 5 vector space or 8 language modelling variants presented in Section \ref{s:ground}) as a proxy of term dependence. This allows to detect queries \textit{more likely to be non-compositional}, hence more likely to contain highly dependent terms, rather than those queries that are \textit{strictly non-compositional}. We do this by ranking queries by their non-compositionality and selecting the $\theta$ least compositional. These $\theta$ queries are processed with the MRF model of fully dependent query terms; 
the rest of the queries in the batch are treated as a bag of words. 

\subsection{Experimental Setup}
\subsubsection{Baselines \& Our Methods} 
We use three baselines: (1) bag of words for all queries, which allows for no term dependence; (2) the MRF model of sequentially dependent query terms \cite{Metzler2005}, which treats as a `set phrase' only adjacent query terms; (3) the MRF model of fully dependent query terms \cite{Metzler2005}, which treats as a `set phrase' the whole query. 
We compare these baselines against our selective term dependence approach that treats as a `set phrase' the whole query iff the NCD score of this query indicates that it is likely to be non-compositional; this is controlled by the threshold $\theta$ presented above.

All three baselines and our 13 NCD variants use a unigram, query likelihood, Dirichlet-smoothed language model for ranking. Note that we use `language model' in two different ways in this work, for two entirely different computations: (a) to estimate the semantic divergence between queries and perturbations (in Section \ref{sss:prob}), and (b) to rank documents with respect to queries.

\subsubsection{Data \& Tuning}
We use the TREC 6-8 queries (301-450, title only) of the AdHoc track with Disks 4-5 (minus the Congressional Records for TREC7-8), and queries 1-200 of the Web AdHoc tracks of TREC 2009-2012 with ClueWeb09B\footnote{http://lemurproject.org/clueweb09.php/} (see Table \ref{tab:datasets}). 
We extract the distributional semantics of the NCD model (i.e. build the context windows) from Disks4-5 for queries 301-450, and from ClueWeb09B for queries 1-200. We use no stemming and remove stop words from the queries only (as in \cite{Metzler2005}).
We use Indri 5.8\footnote{http://www.lemurproject.org/} for indexing and retrieval of at most 1000 documents per query. 
We evaluate retrieval effectiveness using standard measures of early and deep precision (MAP, NDCG@10, P@10). 

The Dirichlet ranking model includes a parameter $\mu$ that we tune as follows: $\mu \in \{100, 500, 800, 1000, 2000, 3000, 4000,$ $ 5000, 8000, 10000\}$. We also vary the number $\theta$ of least compositional queries selected each time: $\theta \in {1 \ldots 45}$ per TREC batch of 50 queries. All tuning is done per evaluation measure using 3-fold cross validation. 
We report the average of the three test folds.  
For NCD we extract the first synonym suggested by WordNet\footnote{http://wordnet.princeton.edu}  (to be used for perturbing the query). 
For these initial experiments, we do not vary the value of the window of co-occurrence described in Section \ref{s:ground}: we set $n=5$, i.e. the context window size is 11. 

\begin{table}
\centering
\caption{\label{tab:datasets}Datasets}
\scalebox{1.0}{
\begin{tabular}{l|c|c} 
			&Disks4-5			&ClueWeb09B\\
\hline
\# Documents	&556077			&50220423\\
\# Queries		&301-450			&1-200\\
TREC track	&TREC6-8 AdHoc	&Web09-12 AdHoc\\
\end{tabular}
}
\centering
\caption{\label{tab:ql}Query length (without stopwords)}
\scalebox{1.0}{
\begin{tabular}{lrrrrr} 
\hline
			&\#1	&\#2	&\#3	&\#4	&\#5\\
\hline
DISKS4-5		&11	&56	&76	&7	&-\\
CWEB09B	&57	&65	&63	&13	&2\\
\hline
\end{tabular}
}
\end{table}


\subsection{Findings}
\label{ss:Results}
\definecolor{light-gray}{gray}{0.90}
\begin{savenotes}
\begin{table*}
\centering
\caption{\label{tab:res1}Retrieval precision of the 3 baselines (in grey rows) vs. our 13 non-compositionality approaches. Bold marks $>$highest baseline. The star * marks best overall per measure \& collection. \%DQ is the \% of queries processed as dependent (the rest of the queries in the batch are processed as bags of words).}
\scalebox{0.82}{
\begin{tabular}{| c | l | l r | l r | l r | l r | l r | l r |}
\hline
&\multirow{2}{*}{METHOD}	&\multicolumn{2}{c|}{DISKS4-5	} &\multicolumn{2}{c|}{CWEB09B}	&\multicolumn{2}{c|}{DISKS4-5	} &\multicolumn{2}{c|}{CWEB09B}	&\multicolumn{2}{c|}{DISKS4-5	} &\multicolumn{2}{c|}{CWEB09B}\\
&			&MAP&\%DQ&MAP&\%DQ&NDCG@10&\%DQ&NDCG10&\%DQ&P@10&\%DQ&P@10&\%DQ\\
\hline
\rowcolor{light-gray}
&Bag of words						&.1905&--		&.1151&--		&.4276&--		&.3502&--			&.3907&--		&.4167&-\\
\rowcolor{light-gray}
&Sequential Dependence	\cite{Metzler2005}	&.1814&100\% &.1077&100\%	&.3983&100\%	&.3463&100\%		&.3687&100\%	&.4120&100\%\\
\rowcolor{light-gray}
&Full Dependence \cite{Metzler2005}		&.1933&100\%	&.1151&100\%	&.4341&100\%	&.3514&100\%		&.4007&100\%	&.4176&100\%\\
\hline
\multirow{8}{*}{\rotatebox[origin=c]{90}{\bf LANG. MODEL}}
&Laplace sum 					&\bf.1948 &63\%		&\bf.1188&34\%	&\bf.4406&70\%	&\bf.3596&18\%		&\bf.4047&67\%	&\bf.4317&30\%\\
&Laplace average 				&\bf.1948&63\%	&\bf.1186&34\%	&\bf.4406&70\%	&\bf.3596&18\%		&\bf.4047&67\%	&\bf.4307&36\%\\
&Laplace median						&\bf.1947&67\%	&\bf.1176&51\%	&\bf.4390&48\%	&\bf.3585&18\%		&\bf.4060&67\%	&\bf.4278&31\%\\
&Laplace multiplication					&\bf.1948&48\%	&\bf.1182&46\%	&\bf.4388&36\%	&\bf.3617&22\%		&\bf.4040&59\%	&\bf.4303&22\%\\
&GoodTuring sum 				&\bf.1940&81\%	&\bf.1168&50\%	&\bf.4402&57\%	&\bf.3618&31\%		&\bf.4040&79\%	&\bf.4288&28\%\\
&GoodTuring average 				&\bf.1940&81\%	&\bf.1167&50\%	&\bf.4402&57\%	&\bf.3618&29\%		&\bf.4040&79\%	&\bf.4288&26\%\\
&GoodTuring median					&\bf.1949&73\%	&\bf.1168&56\%	&\bf.4422&51\%	&\bf.3583&15\%		&\bf.4067*&51\%	&\bf.4283&32\%\\
&GoodTuring multiplication				&\bf.1943&71\%	&\bf.1171&29\%	&\bf.4390&59\%	&\bf.3623&28\%		&\bf.4053&72\%	&\bf.4302&22\%\\
\hline
\multirow{5}{*}{\rotatebox[origin=c]{90}{\bf VECTOR}}
&ATC					&\bf.1950*&77\%				&\bf.1191*&47\%	&\bf.4446*&55\%	&\bf.3604&31\%		&\bf.4053&56\%	&\bf.4308&53\%\\
&LTU					&\bf.1948&75\%				&\bf.1184&40\%	&\bf.4444&51\%	&\bf.3592&29\%		&\bf.4053&52\%	&\bf.4278&33\%\\
&MI						&\bf.1946&81\%				&\bf.1188&51\%	&\bf.4445&59\%	&\bf.3631*&32\%		&\bf.4053&52\%	&\bf.4364*&52\%\\
&Okapi					&\bf.1948&73\%				&\bf.1180&48\%	&\bf.4427&57\%	&\bf.3597&20\%		&\bf.4040&57\%	&\bf.4293&21\%\\
&TFIDF					&\bf.1941&56\%				&\bf.1175&30\%	&\bf.4422&61\%	&\bf.3605&39\%		&\bf.4053&53\%	&\bf.4294&30\%\\	
\hline
\end{tabular}
}
%
\centering
\caption{\label{tab:res2}Retrieval precision for 2/3/4-term queries with our three best non-compositionality approaches. ($\pm$ \%): difference from the strongest baseline. Rest of notation as in Table \ref{tab:res1}.}
\scalebox{0.8}{
\begin{tabular}{ | l |lrr|lrr|lrr| }
\hline
&\multicolumn{9}{c|}{DISKS4-5} \\
METHOD&\multicolumn{3}{c|}{2 terms (56 queries)} &\multicolumn{3}{c|}{3 terms (76 queries)} &\multicolumn{3}{c|}{4 terms (7 queries)}	\\
	&\multicolumn{2}{c}{MAP} &\%DQ&\multicolumn{2}{c}{MAP}&\%DQ&\multicolumn{2}{c}{MAP}&\%DQ\\
\hline
\rowcolor{light-gray}
Bag of words				&.1994&&--		&.1985&&--		&.1181&&--		\\
\rowcolor{light-gray}
Sequential Dependence \cite{Metzler2005}		&.1953&&100\%	&.1722&&100\%	&.1120&&100\%		\\
\rowcolor{light-gray}
Full Dependence \cite{Metzler2005} 			&.2022&&100\%	&.1976&&100\%	&.1143&&100\%		\\
\hline
GoodTuring median		&\bf.2115* &(+4.6\%)&48\%	&\bf.2046* &(+3.1\%)&45\%	&\bf.1245* &(+5.4\%)&29\%	\\
ATC 					&\bf.2114 &(+4.5\%)&48\%	&\bf.2046* &(+3.1\%)&46\%	&\bf.1245* &(+5.4\%)&29\%	\\
MI 						&\bf.2114 &(+4.5\%)&48\%	&\bf.2046* &(+3.1\%)&46\%	&\bf.1245* &(+5.4\%)&29\%	\\
\hline
\hline
&\multicolumn{9}{c|}{DISKS4-5} \\
METHOD&\multicolumn{3}{c|}{2 terms (56 queries)} &\multicolumn{3}{c|}{3 terms (76 queries)} &\multicolumn{3}{c|}{4 terms (7 queries)}	\\
&\multicolumn{2}{c}{NDCG@10} &\%DQ&\multicolumn{2}{c}{NDCG@10}&\%DQ&\multicolumn{2}{c}{NDCG@10}&\%DQ\\
\hline
\rowcolor{light-gray}
Bag of words		&.4331&&--		&.4699&&--		&.3549&&--			\\
\rowcolor{light-gray}
Sequential Dependence \cite{Metzler2005}				&.4183&&100\%	&.3685&&100\%	&.3394&&100\%		\\
\rowcolor{light-gray}
Full Dependence \cite{Metzler2005}			&.4174&&100\%	&.4421&&100\%	&.3768&&100\%		\\
\hline
GoodTuring median		&\bf.4855* &(+12.1\%)&32\%	&\bf.4968* &(+5.7\%)&33\%	&\bf.3902*&(+3.6\%)&29\%	\\
ATC 					&\bf.4855* &(+12.1\%)&32\%	&\bf.4968*&(+5.7\%)&33\%	&\bf.3902*&(+3.6\%)&29\%	\\
MI 						&\bf.4855* &(+12.1\%)&32\%	&\bf.4968*&(+5.7\%)&33\%	&\bf.3902*&(+3.6\%)&29\%	\\
\hline
\hline
&\multicolumn{9}{c|}{DISKS4-5} \\
METHOD&\multicolumn{3}{c|}{2 terms (56 queries)} &\multicolumn{3}{c|}{3 terms (76 queries)} &\multicolumn{3}{c|}{4 terms (7 queries)}	\\
&\multicolumn{2}{c}{P@10} &\%DQ&\multicolumn{2}{c}{P@10}&\%DQ&\multicolumn{2}{c}{P@10}&\%DQ\\\hline
\rowcolor{light-gray}
Bag of words	&.4018&&--		&.4286&&--		&.3000&&--		\\
\rowcolor{light-gray}
Sequential Dependence \cite{Metzler2005}			&.3909&&100\%	&.3429&&100\%	&.3000&&100\%		\\
\rowcolor{light-gray}
Full Dependence \cite{Metzler2005} 		&.3873&&100\%	&.4208&&100\%	&\bf.3400*&&100\%	\\
\hline
GoodTuring median		&\bf.4545*  &(+13.1\%)&20\%	&\bf.4649* &(+8.5\%)&30\%	&\bf.3400* &($\pm$0.0\%)&29\%	\\
ATC 					&\bf.4527 &(+12.7\%)&20\%	&\bf.4649* &(+8.5\%)&30\%	&\bf.3400* &($\pm$0.0\%)&29\%	\\
MI 						&\bf.4527 &(+12.7\%)&20\%	&\bf.4649* &(+8.5\%)&30\%	&\bf.3400* &($\pm$0.0\%)&29\%	\\
\hline
\end{tabular}
}
\scalebox{0.8}{
\begin{tabular}{ | l |lrr|lrr|lrr| }
\hline
&\multicolumn{9}{c|}{CWEB09B}\\
METHOD&\multicolumn{3}{c|}{2 terms (65 queries)} &\multicolumn{3}{c|}{3 terms (63 queries)} &\multicolumn{3}{c|}{4 terms (13 queries)}\\
	&\multicolumn{2}{c}{MAP} &\%DQ&\multicolumn{2}{c}{MAP}&\%DQ&\multicolumn{2}{c}{MAP}&\%DQ\\
	\hline
\rowcolor{light-gray}
Bag of words					&.1290&&--		&.1391&&--		&.1046&&--\\
\rowcolor{light-gray}
Sequential Dependence \cite{Metzler2005}			&.1126&&100\%	&.1235&&100\%	&.0949&&100\%\\
\rowcolor{light-gray}
Full Dependence \cite{Metzler2005} 					&.1234&&100\%	&.1377&&100\%	&.0982&&100\%\\
\hline
GoodTuring median			&\bf.1371 &(+6.3\%)&43\%	&\bf.1480 &(+6.4\%)&43\%	&\bf.1120 &(+7.1\%)&15\%\\
ATC 						&\bf.1368 &(+6.0\%)&48\%	&\bf.1519 &(+9.2\%)&46\%	&\bf.1128* &(+7.8\%)&23\%\\
MI 							&\bf.1368 &(+6.0\%)&48\%	&\bf.1520* &(+9.3\%)&46\%	&\bf.1128* &(+7.8\%)&23\%\\
\hline
\hline
&\multicolumn{9}{c|}{CWEB09B}\\
METHOD&\multicolumn{3}{c|}{2 terms (65 queries)} &\multicolumn{3}{c|}{3 terms (63 queries)} &\multicolumn{3}{c|}{4 terms (13 queries)}\\
&\multicolumn{2}{c}{NDCG@10} &\%DQ&\multicolumn{2}{c}{NDCG@10}&\%DQ&\multicolumn{2}{c}{NDCG@10}&\%DQ\\
\hline
\rowcolor{light-gray}
Bag of words			&.4003&&--		&.2907&&--		&.3552&&--	\\
\rowcolor{light-gray}
Sequential Dependence \cite{Metzler2005}					&.3671&&100\%	&.2902&&100\%	&.2409&&100\%\\
\rowcolor{light-gray}
Full Dependence \cite{Metzler2005}					&.3412&&100\%	&.2958&&100\%	&.3213&&100\%\\
\hline
GoodTuring median			&\bf.4142 &(+3.5\%)&32\%	&\bf.3267  &(+10.4\%)&33\%	&\bf.3873*  &(+9.8\%)&31\%\\
ATC 					&\bf.4143  &(+3.5\%)&34\%	&\bf.3291*  &(+11.3\%)&35\%	&\bf.3838  &(+8.1\%)&31\%\\
MI 						&\bf.4142  &(+3.5\%)&34\%	&\bf.3291*  &(+11.3\%)&35\%	&\bf.3838  &(+8.1\%)&31\%\\
\hline
\hline
&\multicolumn{9}{c|}{CWEB09B}\\
METHOD&\multicolumn{3}{c|}{2 terms (65 queries)} &\multicolumn{3}{c|}{3 terms (63 queries)} &\multicolumn{3}{c|}{4 terms (13 queries)}\\
&\multicolumn{2}{c}{P@10} &\%DQ&\multicolumn{2}{c}{P@10}&\%DQ&\multicolumn{2}{c}{P@10}&\%DQ\\\hline
\rowcolor{light-gray}
Bag of words		&.4894&&--		&.3600&&--		&.3538&&-\\
\rowcolor{light-gray}
Sequential Dependence \cite{Metzler2005}					&.4318&&100\%	&.3550&&100\%	&.2692&&100\%\\
\rowcolor{light-gray}
Full Dependence \cite{Metzler2005} 			&.4167&&100\%	&.3617&&100\%	&.3615&&100\%\\
\hline
GoodTuring median			&\bf.5152 &(+5.3\%)&20\%	&\bf.4067  &(+12.4\%)&21\%	&\bf.4154*  &(+14.9\%)&38\%\\
ATC 						&\bf.5167*  &(+5.6\%)&25\%	&\bf.4100*  &(+13.4\%)&19\%	&\bf.4154*  &(+14.9\%)&38\%\\
MI 						&\bf.5167*  &(+5.6\%)&25\%	&\bf.4100*  &(+13.4\%)&19\%	&\bf.4154*  &(+14.9\%)&38\%\\
\hline
\end{tabular}
}

\end{table*}
\end{savenotes}

Table \ref{tab:res1} shows the retrieval precision of our baselines and NCD approaches. 
Each cell also displays the \% of queries that are treated as a `set phrase'. 
For the MRF models, 100\% means that all queries are treated as a `set phrase', including single-term queries, for which this treatment makes no difference over a bag of words treatment. Overall, our NCD approaches outperform all baselines at all times. The improvement over the strongest baseline is modest (up to >$+3.5\%$ for MAP with ATC, >$+3.3\%$ for NDCG@10 with MI, and >$+4.5\%$ for P@10 with MI), however it is consistent for both datasets and for all evaluation measures (deep and early precision). This means that the performance gain spans across the range of relevant documents (those retrieved in the top ranks, but also those retrieved further down). Unlike earlier findings that the use of co-occurrence information tends to reduce retrieval effectiveness \cite{Salton1982}, possibly due to the fact that the term relationships modelled may have little discriminating power \cite{Metzler2005}, we notice an overall modest but clear gain in effectiveness. 

Breaking this down to a per-query basis (cf. the two top plots in Fig. \ref{fig:diff}), the following two findings emerge. (I) The scale of improvement is higher than that of deterioration: between $\sim$+0.13 and -0.07 for MAP; and between +0.68 and -0.4 for NDCG@10, for our Laplace sum approach (chosen illustritatively) from the strongest baseline (MRF with full dependence). (II) More queries improve than deteriorate by our approach. Hence, the improvements in Table \ref{tab:res1} are not artificially inflated by outliers that might affect the means of the evaluation measures, but are rather representative of the whole body of queries. 

Furthermore, we show examples of queries yielding the highest and lowest precision difference from the strongest baseline in Table \ref{tab:best}. The best queries are \textit{not} strictly non-compositional; however they do have strongly contextualised semantics and term co-dependence. E.g. \texttt{french lick resort casino} does not denote some other meaning than a particular casino, but it is presumably irrelevant to the semantics of the verb \texttt{to lick} and \texttt{french} as a language or nationality. Most of the best queries in Table \ref{tab:best} are web queries, which often tend to include abbreviations and acronyms, e.g. \texttt{vbart sf}. These are not non-compositional either, but rather idiomatic or colloquial phrases of strong term dependence, and are selected by our NCD approach because they are likely to diverge in meaning if perturbed (i.e. it is not possible to express their meaning alternatively, for instance by near-synonyms). 
Hence, using NCD to approximate strong term dependence is effective in these cases.
Our worst performing queries consist of phrases for which many more variants that denote the same meaning exist. E.g. \texttt{tv show, television programme/broadcast, signs/symptoms/indications heart attack/failure}, etc. Restricting this type of queries to strict `set phrase' matching limits the retrieval scope significantly with resulting drops in performance. 


Next we focus the analysis on two pertinent aspects of our approach: the number of strongly term dependent queries selected and retrieval performance for 2-4 term queries.

\subsubsection{Number of least compositional queries}
The number of queries treated as a `set phrase' is lower for our approach than for MRF by $\sim$1/3 for Disks4-5 and $\sim$2/3 for ClueWeb09B, or by $\sim$1/4 for Disks4-5 and $\sim$1/3 for ClueWeb09B if we ignore 1-term queries 
(statistics in Table \ref{tab:ql}). Compositionality and term dependence in general cannot be measured for single terms, hence 1-term queries are ignored. 

Since we treat the number $\theta$ of least compositional queries as a tuneable parameter, one may wonder to what extent the gains we report are due to tuning as opposed to the inherent strength of our approach in detecting term dependence. To answer this, Fig. \ref{fig:k} shows the MAP and NDCG@10 of our MI approach across the range of $\theta$ values for ClueWeb09B (we can confirm similar trends for P@10 and Disks4-5, and our other NCD approaches). We see that our approach outperforms the strongest baseline (marked by a horizontal line) consistently across the range of $\theta$, peaking when roughly $\theta=$80 least compositional queries (out of 200, or 143 if one excludes 1-term queries) are treated as strongly term dependent. Practically this means that our approach can be used without necessarily tuning $\theta$ and is likely not to give large fluctuations in both early and deep precision.

\subsubsection{Queries of 2-4 terms}
Finally, we focus on queries of 2, 3 and 4 terms because these are the most likely to include strong term dependence, hence they are ideal for comparing our approaches to the MRF models.

Table \ref{tab:res2} shows the retrieval precision of our baselines and our three best NCD approaches (marked by * in Table \ref{tab:res1}) specifically for queries of these lengths. Again all our approaches outperform all baselines at all times. The only exception is for 4-term queries in Disks4-5 and P@10, where our methods perform equally to the strongest baseline (no gain, no loss). Overall, our NCD approaches outperform the strongest baseline by up to $\sim$>$+5\%$ for MAP,  $\sim$>$+6\%$ for NDCG@10, and $\sim$>$+8\%$ for P@10, on average.  The two middle and lower plots in Fig. \ref{fig:diff} show that these improvements are not due to outliers, but are instead spread over the queries. Fig. \ref{fig:diff} illustrates this for 2- and 3-term queries w.r.t. MAP and NDCG@10, but we confirm that the same trend applies to 4-term queries and P@10. Hence, for queries of length 2-4, i.e. predominantly phrasal queries, our approaches outperform all baselines notably. This finding, combined with the relative robustness of the threshold $\theta$ discussed above, mean that our approach could be used as part of the IR pipeline, e.g. for $\sim$80\% of the incoming queries of length 2-4. Note that these types of queries form the majority of all queries, at least in our TREC data (see Table \ref{tab:ql}), hence they are not a negligible sample.



\begin{figure}
\centering
\pgfplotsset{every axis label/.append style={font=\tiny}}
\tikzset{every mark/.append style={font=\tiny}}
\pgfplotsset{every axis legend/.append style={
			at={(0.5,-0.0)},
			anchor=south}
			}
\scalebox{0.78}{
\begin{tabular}{cc}
\begin{tikzpicture}[baseline,scale=0.5]
\begin{axis}[
scale only axis,
hide x axis=true,
font=\LARGE,
scaled y ticks = false,
y tick label style={/pgf/number format/fixed},
title=\textbf{MAP all 350 queries}
]
\pgfplotstableread{fd-nc-map.plot}\table 
\addplot+[mark=+,only marks,blue] table[x index=0,y index=1] from \table;
\addplot+[smooth,black] coordinates{(0,0) (350,0)};
\end{axis}
\end{tikzpicture} &
\begin{tikzpicture}[baseline,scale=0.5]
\begin{axis}[
scale only axis,
hide x axis=true,
font=\LARGE,
scaled y ticks = false,
y tick label style={/pgf/number format/fixed},
title=\textbf{NDCG@10 all 350 queries}
]
\pgfplotstableread{fd-nc-ndcg.plot}\table 
\addplot+[mark=+,only marks,blue] table[x index=0,y index=1] from \table;
\addplot+[smooth,black] coordinates{(0,0) (350,0)};
\end{axis}
\end{tikzpicture}\\
\begin{tikzpicture}[baseline,scale=0.5]
\begin{axis}[
scale only axis,
hide x axis=true,
font=\LARGE,
scaled y ticks = false,
y tick label style={/pgf/number format/fixed},
title=\textbf{MAP 121 2-term queries }
]
\pgfplotstableread{fd-nc-map-2.plot}\table 
\addplot+[mark=+,only marks,blue] table[x index=0,y index=1] from \table;
\addplot+[smooth,black] coordinates{(0,0) (121,0)};
\end{axis}
\end{tikzpicture} &
\begin{tikzpicture}[baseline,scale=0.5]
\begin{axis}[
scale only axis,
hide x axis=true,
font=\LARGE,
scaled y ticks = false,
y tick label style={/pgf/number format/fixed},
title=\textbf{NDCG@10 121 2-term queries}
]
\pgfplotstableread{fd-nc-ndcg-2.plot}\table 
\addplot+[mark=+,only marks,blue] table[x index=0,y index=1] from \table;
\addplot+[smooth,black] coordinates{(0,0) (121,0)};
\end{axis}
\end{tikzpicture}\\
\begin{tikzpicture}[baseline,scale=0.5]
\begin{axis}[
scale only axis,
hide x axis=true,
font=\LARGE,
scaled y ticks = false,
y tick label style={/pgf/number format/fixed},
title=\textbf{MAP 139 3-term queries}
]
\pgfplotstableread{fd-nc-map-3.plot}\table 
\addplot+[mark=+,only marks,blue] table[x index=0,y index=1] from \table;
\addplot+[smooth,black] coordinates{(0,0) (139,0)};
\end{axis}
\end{tikzpicture} &
\begin{tikzpicture}[baseline,scale=0.5]
\begin{axis}[
scale only axis,
hide x axis=true,
font=\LARGE,
scaled y ticks = false,
y tick label style={/pgf/number format/fixed},
title=\textbf{NDCG@10 139 3-term queries}
]
\pgfplotstableread{fd-nc-ndcg-3.plot}\table 
\addplot+[mark=+,only marks,blue] table[x index=0,y index=1] from \table;
\addplot+[smooth,black] coordinates{(0,0) (139,0)};
\end{axis}
\end{tikzpicture}
\end{tabular}
}
\caption{\label{fig:diff} Sorted per-query difference (y-axis) in MAP/NDCG@10 between the strongest baseline (\textit{Full Dependence}) and our \textit{Laplace sum} method, for all, 2-term, \& 3-term queries in DISKS4-5 \& CWEB09B. The horizontal line marks the baseline (points above are gains). Each point is a query.}
\end{figure}
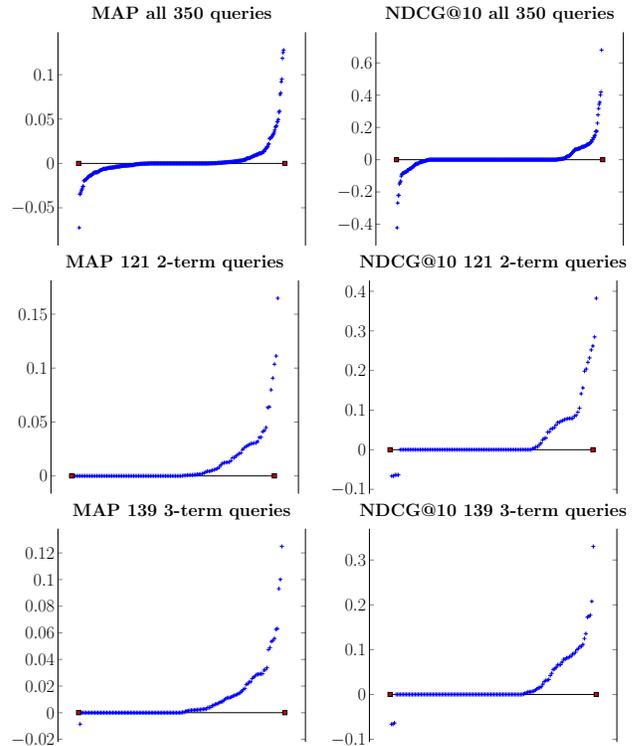

\begin{figure}
\centering
\pgfplotsset{every axis label/.append style={font=\tiny}}
\tikzset{every mark/.append style={font=\tiny}}
\pgfplotsset{every axis legend/.append style={
			at={(0.5,-0.0)},
			anchor=south}
			}
\scalebox{0.87}{
\begin{tabular}{cc}
\begin{tikzpicture}[baseline,scale=0.5]
\begin{axis}[
ytick={0.1151,0.1188},
font=\LARGE,
y tick label style={/pgf/number format/precision=5},
title=\textbf{MAP}
]
\pgfplotstableread{wt-MI-map-theta.plot2}\table 
\addplot+[mark=+,only marks,blue] table[x index=0,y index=1] from \table;
\addplot+[smooth,black] coordinates{(0,0.1151) (120,0.1151)};
\end{axis}
\end{tikzpicture}
\begin{tikzpicture}[baseline,scale=0.5]
\begin{axis}[
ytick={0.3514,0.3600},
font=\LARGE,
scaled y ticks = false,
y tick label style={/pgf/number format/precision=5},
title=\textbf{NDCG@10}
]
\pgfplotstableread{wt-MI-ndcg_cut_10-theta.plot2}\table 
\addplot+[mark=+,only marks,blue] table[x index=0,y index=1] from \table;
\addplot+[smooth,black] coordinates{(0,0.3514) (120,0.3514)};
\end{axis}
\end{tikzpicture}
\end{tabular}
}
\caption{\label{fig:k} MAP \& NDCG@10 (y-axis) vs. $\theta$ most non-compositional  queries in CWEB09B according to MI (x-axis). The horizontal line marks the baseline. Each point is a query.}
\end{figure}
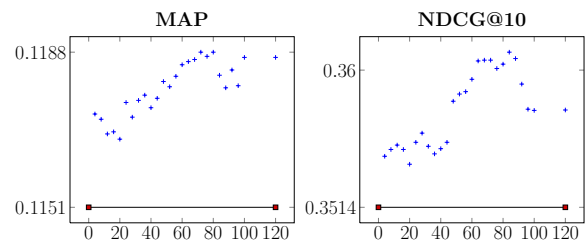
\begin{table}
\centering
\caption{\label{tab:best}Queries with most gain/loss from NCD.}
\scalebox{0.87}{
\begin{tabular}{c|c} 
\hline
Best&Worst\\
\hline
\textit{bart sf}				&\textit{tv show}\\
\textit{ct jobs}				&\textit{industrial espionage}\\
\textit{french lick resort casino}	&\textit{export controls cryptography}	\\
\textit{civil right movement}	&\textit{signs heartattack}\\
\hline
\end{tabular}
}
\end{table}

\section{Discussion}
\label{s:Discussion}

The relative gains in retrieval precision reported above should not be considered as indications of accurate non-compositionality detection. The suitability of our proposed probabilistic representation of queries and their perturbations in particular remains to be evaluated for NCD accuracy. Moreover, several of our choices of NCD settings can be further explored, e.g. synonymy selection or smoothing choices. In this initial study we opted for default or popular settings, where possible. For these reasons, we have refrained from making a quantitative comparison between the vector space and probability space NCD variations, other than reporting the retrieval precision they yield. This means that the NCD variations we present are not necessarily calibrated to this domain or task. Calibrating them could potentially improve performance even more, but would incur some computational cost, the major bulk of which would likely lie in the extraction of context windows from some large dataset. In an IR scenario, this can be done offline, and is perhaps not too distant from the query analytics widely used. 



Regarding our data, the query sets we use are `curated' by TREC, in the sense that those queries that are perhaps not understood by human assessors, or for which no relevant documents are easily found during pooling, may have been omitted. This selection may have affected non- or low-compositionality queries. This agrees with the finding that the number of IR benchmark queries that contain strongly dependent terms in general is small \cite{ZobelM2006}. Unfiltered query logs may contain more such queries, making our approach potentially even more useful in such a practical setting.




\section{Conclusions}
\label{s:Conclusions}
We presented an approach for detecting strongly dependent query terms using the linguistic property of non-compo-\\sitionality. Non-compositional meaning cannot be induced from the meanings of individual words or their arrangement in a query. E.g., \texttt{hot dog} is not a type of \texttt{dog} that is \texttt{hot}, but rather a type of food. We used unsupervised measurement of non-compositionality to approximate the detection of strongly dependent query terms. Such queries are challenging to IR because they cannot be processed to some reasonable accuracy by bag of words approaches. Motivated by this, we focussed not on how these queries can be treated during ranking (there is a lot of literature in this area generally for term dependence, which can be applied here), but on how these queries can be selected from a batch or stream of incoming queries. This specific question has so far been addressed by assuming that the more frequently terms co-occur in a query, the more dependent they are. This assumption is however not always true, because frequency is not always proportional to the strength of semantic association. The unsupervised method for measuring non-compositionality that we used is recent and uses vector spaces \cite{KielaC13}. We extended it by adding a probabilistic representation that uses Kullback-Leibler divergence. We experimentally showed that  all variants of our approach were effective in selecting which queries to treat as term dependent and resulted in gains for both early and deep precision ($>5\%$) with respect to a range of baselines (standard bag of words and competitive MRF with sequential and full dependence \cite{Metzler2005}). 

In the future we plan to analyse the amount of non- or low-compositionality queries in real-life query logs, as opposed to TREC data. As discussed in Section \ref{s:Discussion}, there may be more low-compositionality queries in those samples. We also intend to investigate optimal ways of measuring non-compositionality \textit{within} a query, as opposed to considering the non-compositionality of a query as a whole as we did here. Another interesting direction is the direct mapping of the non-compositionality score of a query into the strength of its term dependence used during ranking. In this initial study we treated all queries selected as least-compositional in the same way as fixed phrases processing them identically; in doing so, we ignored their grades of non-compositionality. Modelling this may yield further improvements and is an interesting research question in its own right.

\paragraph{Acknowledgments}
Partially funded by the first author's \textit{FREJA research excellence} fellowship (grant no. 790095).

%

\begin{thebibliography}{10}

\bibitem{Baldwin:2003}
T.~Baldwin, C.~Bannard, T.~Tanaka, and D.~Widdows.
\newblock An empirical model of multiword expression decomposability.
\newblock In {\em ACL Multiword Expressions Wksh.}, pages 89--96. 2003.

\bibitem{bannard2003statistical}
C.~Bannard, T.~Baldwin, and A.~Lascarides.
\newblock A statistical approach to the semantics of verb-particles.
\newblock In {\em ACL Multiword Expressions Wksh.}, pages 65--72, 2003.

\bibitem{Baxendale1958}
P.~B. Baxendale.
\newblock Machine-made index for technical literature.
\newblock {\em IBM Journal for R\&D}, 2:354--361, 1958.

\bibitem{BenderskyMC10}
M.~Bendersky, D.~Metzler, and W.~B. Croft.
\newblock Learning concept importance using a weighted dependence model.
\newblock In {\em WSDM}, pages 31--40, 2010.

\bibitem{DillonG1983}
M.~Dillon and A.~Gray.
\newblock {FASIT} - a fully automatic syntactically based indexing system.
\newblock {\em JASIS}, 34:99--108, 1983.

\bibitem{Doyle1962}
L.~B. Doyle.
\newblock Indexing and abstracting by association. part i.
\newblock {\em Am. Doc.}, 13:378--390, 1962.

\bibitem{Earl1972}
L.~L. Earl.
\newblock The resolution of syntactic ambiguity in automatic language
  processing.
\newblock {\em Information Storage and Retrieval}, 8:277--308, 1972.

\bibitem{LNC3:LNC3362}
K.~Erk.
\newblock Vector space models of word meaning and phrase meaning: A survey.
\newblock {\em Language and Linguistics Compass}, 6(10):635--653, 2012.

\bibitem{EvansZ1996}
D.~A. Evans and C.~Zhai.
\newblock Noun-phrase analysis in unrestricted text for {IR}.
\newblock In {\em ACL}, pages 17--24, 1996.

\bibitem{Fagan1989}
J.~L. Fagan.
\newblock The effectiveness of a non syntactic approach to automatic phrase
  inducing for document retrieval.
\newblock {\em JASIS}, 40:115--132, 1989.

\bibitem{firth}
J.~R. Firth.
\newblock A synopsis of linguistic theory.
\newblock {\em Selected papers of J.R. Firth 1952-1959}, pages 168--205, 1968.

\bibitem{Fujita2001}
S.~Fujita.
\newblock More reflections on aboutness {TREC-2001} evaluation experiments at
  {Justsystem}.
\newblock In {\em TREC}, pages 331 -- 338, 2001.

\bibitem{gale1995good}
W.~Gale and G.~Sampson.
\newblock {Good-Turing} frequency estimation without tears.
\newblock {\em J. of Quant. Ling.}, 2(3):217--237, 1995.

\bibitem{Giuliano1963}
V.~E. Giuliano and P.~E. Jones.
\newblock Linear associative ir.
\newblock {\em Vistas in Information Handling: The Augmentation of Man's
  Intellect by Machine}, 1:30--54, 1963.

\bibitem{HarperK1978}
D.~J. Harper and C.~J.~K. van Rijsbergen.
\newblock An evaluation of feedback in document retrieval using concurrence
  data.
\newblock {\em J. of Doc.}, 34:189--216, 1978.

\bibitem{Hey2002}
F.~Heylighen and J.~Dewaele.
\newblock Variation in the contextuality of language.
\newblock {\em F. of Sci.}, 7(3):293--340, 2002.

\bibitem{Jeffreys48}
H.~Jeffreys.
\newblock {\em Theory of Probability}.
\newblock Clarendon, 1948.

\bibitem{JinFH01}
R.~Jin, C.~Falusos, and A.~G. Hauptmann.
\newblock Meta-scoring: Automatically evaluating term weighting schemes in {IR}
  without precision-recall.
\newblock In {\em SIGIR}, pages 83--89, 2001.

\bibitem{JurafskyM09}
D.~Jurafsky and J.~Martin.
\newblock {\em Speech and Language Processing}.
\newblock Pearson, 2009.

\bibitem{Katz:automatic}
G.~Katz and E.~Giesbrecht.
\newblock Automatic identification of non-compositional multi-word expressions
  using {LSA}.
\newblock In {\em ACL Multiword Expressions Wksh.}, pages 12--19. 2006.

\bibitem{KielaC13}
D.~Kiela and S.~Clark.
\newblock Detecting compositionality of multi-word expressions using nearest
  neighbours in vector space models.
\newblock In {\em EMNLP}, pages 1427--1432, 2013.

\bibitem{Korkontzelos:2009:DCM:1667583.1667605}
I.~Korkontzelos and S.~Manandhar.
\newblock Detecting compositionality in multi-word expressions.
\newblock In {\em ACL-IJCNLP}, pages 65--68, 2009.

\bibitem{krvcmavr-jevzek-pecina:2013:CVSC}
L.~Kr\v{c}m\'{a}\v{r}, K.~Je\v{z}ek, and P.~Pecina.
\newblock Determining compositionality of expresssions using various word space
  models and methods.
\newblock In {\em Continuous Vector Space Models and their Compositionality
  Wksh.}, pages 64--73, 2013.

\bibitem{Lesk1969}
M.~E. Lesk.
\newblock Word-word associations in document retrieval systems.
\newblock {\em Am. Doc.}, 20:27--38, 1969.

\bibitem{Lewis1992}
D.~D. Lewis.
\newblock An evaluation of phrasal and clustered representations on a text
  categorization task.
\newblock In {\em SIGIR}, pages 37--50, 1992.

\bibitem{LewisC1990}
D.~D. Lewis and W.~B. Croft.
\newblock Term clustering of syntactic phrases.
\newblock In {\em SIGIR}, pages 385--404, 1990.

\bibitem{Lin2001}
J.~Lin.
\newblock {Indexing \& Retrieving Natural Language Using Ternary Expressions}.
\newblock Master's thesis, U. of Maryland, USA, 2001.

\bibitem{Losee1994}
R.~Losee.
\newblock Term dependence: Truncating the {Bahadur Lazarsfeld} expansion.
\newblock {\em IPM}, 30(2):293--303, 1994.

\bibitem{LvZ09}
Y.~Lv and C.~Zhai.
\newblock Positional language models for information retrieval.
\newblock In {\em SIGIR}, pages 299--306, 2009.

\bibitem{McCarthy:2003}
D.~McCarthy, B.~Keller, and J.~Caroll.
\newblock Detecting a continuum of compositionality in phrasal verbs.
\newblock In {\em ACL Multiword Expressions Wksh.}, pages 73--80. 2003.

\bibitem{Metzler2005}
D.~Metzler and B.~Croft.
\newblock A {MRF} model for term dependencies.
\newblock In {\em SIGIR}, pages 472--479, 2005.

\bibitem{Metzler1984}
D.~P. Metzler, T.~Noreault, L.~Richey, and P.~B. Heidorn.
\newblock Dependency parsing for information retrieval.
\newblock In {\em SIGIR}, pages 313--324, 1984.

\bibitem{mich11}
L.~Michelbacher, A.~Kothari, M.~Forst, C.~Lioma, and H.~Sch{\"u}tze.
\newblock A cascaded classification approach to semantic head recognition.
\newblock In {\em EMNLP}, pages 793--803, 2011.

\bibitem{MishneR2005}
G.~Mishne and M.~de~Rijke.
\newblock Boosting web retrieval through query operations.
\newblock In {\em ECIR}, pages 502--516, 2005.

\bibitem{COGS:COGS1106}
J.~Mitchell and M.~Lapata.
\newblock Composition in distributional models of semantics.
\newblock {\em Cognitive Science}, 34(8):1388--1429, 2010.

\bibitem{NallapatiA2002}
R.~Nallapati and J.~Allan.
\newblock Capturing term dependencies using a language model based on sentence
  trees.
\newblock In {\em CIKM}, pages 383--390, 2002.

\bibitem{NaritaO2000}
M.~Narita and Y.~Ogawa.
\newblock The use of phrases from query texts in {IR}.
\newblock In {\em SIGIR}, pages 318--320, 2000.

\bibitem{PantelL02}
P.~Pantel and D.~Lin.
\newblock Document clustering with committees.
\newblock In {\em SIGIR}, pages 199--206. ACM, 2002.

\bibitem{Pederson1997}
J.~Pederson, C.~Silverstein, and C.~Vogt.
\newblock Verity at {TREC-6}: out-of-the-box and beyond.
\newblock In {\em TREC-6}, pages 259 -- 274, 1997.

\bibitem{PlachourasO2007}
V.~Plachouras and I.~Ounis.
\newblock Multinomial randomness models for retrieval with document fields.
\newblock In {\em ECIR}, pages 28--39, 2007.

\bibitem{DBLP:conf/ijcnlp/ReddyKMM11}
S.~Reddy, I.~Klapaftis, D.~McCarthy, and S.~Manandhar.
\newblock Dynamic \& static prototype vectors for semantic composition.
\newblock In {\em IJCNLP}, pages 705--713, 2011.

\bibitem{McCarthy:2011}
S.~Reddy, D.~McCarthy, S.~Manandhar, and S.~Gella.
\newblock Exemplar-based word-space model for compositionality detection:
  shared task system description.
\newblock In {\em DiSCo}, pages 54--60. 2003.

\bibitem{Reed2006}
J.~W. Reed, Y.~Jiao, T.~E. Potok, B.~A. Klump, M.~T. Elmore, and A.~R. Hurson.
\newblock {TF-ICF}: A new term weighting scheme for clustering dynamic data
  streams.
\newblock In {\em ICMLA}, pages 258--263, 2006.

\bibitem{Salehi:2013}
B.~Salehi and P.~Cook.
\newblock Predicting the compositionality of multiword expressions using
  translations in multiple languages.
\newblock In {\em SEM}, pages 266--275. 2013.

\bibitem{Salton1966}
G.~Salton.
\newblock Automatic phrase matching.
\newblock {\em Readings in Automatic Language Processing}, pages 169--188,
  1966.

\bibitem{Salton1982}
G.~Salton, C.~Buckley, and C.~T. Yu.
\newblock An evaluation of term dependence models in information retrieval.
\newblock In {\em SIGIR}, pages 151--173, 1982.

\bibitem{Schulte:2013}
S.~Schulte~im Walde, S.~M{\"u}ller, and S.~Roller.
\newblock Exploring vector space models to predict the compositionality of
  {G}erman noun-noun compounds.
\newblock In {\em SEM}, pages 255--265. 2013.

\bibitem{Singhal97}
A.~Singhal.
\newblock {AT{\&}T} at {TREC-6}.
\newblock In {\em TREC-6}, pages 215--225, 1997.

\bibitem{SmeatonK1988}
A.~Smeaton and K.~van Rijsbergen.
\newblock Experiment on incorporation syntactic processing of user queries into
  a document retrieval strategy.
\newblock In {\em SIGIR}, pages 31--51, 1988.

\bibitem{Smeaton1986}
A.~F. Smeaton.
\newblock Incorporating syntactic information into a document retrieval
  strategy: An investigation.
\newblock In {\em SIGIR}, pages 103--113, 1986.

\bibitem{SmithD1985}
F.~J. Smith and K.~Devine.
\newblock Storing and retrieving word phrases.
\newblock {\em IPM}, 21(3):215--224, 1985.

\bibitem{SongC1999}
F.~Song and W.~B. Croft.
\newblock A general language model for {IR}.
\newblock In {\em CIKM}, pages 316--321, 1999.

\bibitem{ksj72}
K.~Sp{\"a}rck-Jones.
\newblock A statistical interpretation of term specificity and its application
  in retrieval.
\newblock {\em J. of Doc.}, 28(1):132--142, 1972.

\bibitem{jones2005}
K.~Sp{\"a}rck-Jones and J.~Tait.
\newblock {\em Charting a New Course: MLP and IR: Essays in Honour of Karen
  Sp{\"a}rck Jones}.
\newblock Springer, 2005.

\bibitem{SrikanthS03}
M.~Srikanth and R.~K. Srihari.
\newblock Incorporating query term dependencies in language models for document
  retrieval.
\newblock In {\em SIGIR}, pages 405--406. ACM, 2003.

\bibitem{Stiles1961}
H.~E. Stiles.
\newblock The association factor in information retrieval.
\newblock {\em Journal of the ACM}, 8:271--279, 1961.

\bibitem{Str1997}
T.~Strzalkowski and F.~Lin.
\newblock Natural language {IR TREC-6} report.
\newblock In {\em TREC}, pages 347 -- 366, 1997.

\bibitem{TaoZ07}
T.~Tao and C.~Zhai.
\newblock An exploration of proximity measures in ir.
\newblock In W.~Kraaij, A.~P. de~Vries, C.~L.~A. Clarke, N.~Fuhr, and N.~Kando,
  editors, {\em SIGIR}, pages 295--302. ACM, 2007.

\bibitem{frege}
R.~H. Thomason.
\newblock {\em Formal Philosophy. Selected Papers of Richard Montague}.
\newblock Yale University Press, 1974.

\bibitem{Turtle1991}
H.~Turtle and B.~Croft.
\newblock Evaluation of an inference network-based retrieval model.
\newblock {\em TOIS}, 9(3):187--222, 1991.

\bibitem{K1977}
C.~J.~K. van Rijsbergen.
\newblock A theoretical basis for the use of co-occurrence data in information
  retrieval.
\newblock {\em J. Doc.}, 33:106--119, 1977.

\bibitem{VinayCMW06}
V.~Vinay, I.~J. Cox, N.~Milic-Frayling, and K.~R. Wood.
\newblock On ranking the effectiveness of searches.
\newblock In {\em SIGIR}, pages 398--404, 2006.

\bibitem{Yu1983}
C.~T. Yu, C.~Buckley, K.~Lam, and G.~Salton.
\newblock A generalised term dependence model in {IR}.
\newblock {\em Information Technology: R\&D}, 2:129--154, 1983.

\bibitem{Zhai1997}
C.~Zhai, X.~Tong, N.~Milic-frayling, and D.~A. Evans.
\newblock Evaluation of syntactic phrase indexing - {CLARIT NLP} track report.
\newblock In {\em TREC-5}, pages 347--358, 1997.

\bibitem{ZobelM2006}
J.~Zobel and A.~Moffat.
\newblock Inverted files for text search engines.
\newblock {\em ACM Comput. Surv.}, 38(2):6, 2006.

\end{thebibliography}

%
%

\end{document}